# High-power magnetron transmitter as an RF source for the electron collider ring of the MEIC facility.

G. Kazakevich, Muons, Inc.

We propose a novel concept of high-power transmitters utilizing the Continuous Wave (CW) magnetrons, injection-locked by phase-modulated signals for compensating of energy losses caused by Synchrotron Radiation (SR) in the electron collider ring of the MEIC facility. At operating frequency of about 750 MHz the SR losses are ~2 MW, [1, 2]. They can be compensated by some number of Superconducting RF (SRF) cavities at a feeding power of about 100-200 kW per cavity.

The investment costs for an RF power system for this task is a notable fraction of the overall costs, if traditional RF sources as klystrons, Inductive Output Tubes (IOTs) or solid-state amplifiers are used. Utilization of MW-scale CW klystrons to power groups of the cavities can save costs to some extent, but in turn only allows a control of the vector sum of the accelerating voltage in the group. Accelerating voltage vector sum control may be undesirable since non-optimized values of phase and amplitude of the accelerating field in individual SRF cavities can cause emittance growth, and may result in droop of luminosity in the collider.

The CW magnetrons based on commercial prototypes are more efficient and potentially less expensive than the above-mentioned RF sources, thus utilization of the magnetron RF sources in the large-scale accelerator projects will notably reduce the capital and maintenance costs, especially since the CW magnetrons with power up to 100-150 kW are well within current manufacturing capabilities.

Unlike requirements for feeding of room-temperature electron accelerating structures, where frequency and phase stabilities of the magnetron RF sources are general tasks, feeding of the SRF cavities requires first of all high stability in phase and amplitude of the accelerated field in the cavity, since some parasitic modulations of the accelerating field are not associated with stability of the feeding RF source. E.g. "microphonics", dynamic tuning errors, beam loading cause the parasitic modulations even the RF source is ideally stable. Thus a traditional concept of the magnetrons, injection-locked in frequency and phase is not acceptable for feeding of the SRF cavities. Instead of this we propose to power the SRF cavities of the electron collider ring by a magnetron transmitters based on the CW magnetrons, injection (frequency)-locked by the wide-band phase-modulated signal, [3, 4].

The wide-band phase modulation of the injection-locked signal will dynamically stabilize phase and amplitude of the accelerated field in a SRF cavity by a Low Level RF (LLRF) system associated with a wide-band closed feedback loop. Emulation of the proposed concept based on the experimental tests demonstrates elimination of low-frequency parasitic modulation caused by "microphonics", dynamic tuning errors, beam loading, and own perturbations of the magnetron phase, caused by phase pushing, distortions of magnetic field, etc., [3].

Proof of principle of the proposed magnetron transmitter concept and concept of injection-locking of magnetrons by phase-modulated signals, realizing a wide-band dynamic control of the accelerating field in SRF cavities were demonstrated with 2.45 GHz, CW, 1 kW magnetrons.

The conceptual scheme of the high-power CW magnetron transmitter allowing the frequency-locking by the phase-modulated signal is shown in Fig. 1. Proof of principle of the transmitter concept was demonstrated in experiments conducted with CW 2.45 GHz, 1 kW magnetrons, operating in general in pulsed mode at pulse duration of 5 ms. Two magnetrons with close free run frequencies were chosen for the tests to be locked at the same frequency. The both magnetrons were mounted in separate modules, Fig. 2, allowing to measure phase performance of the injection-locked magnetrons at various setups.

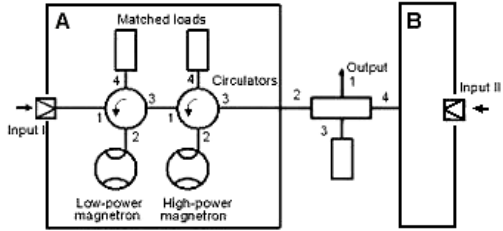

Fig.1: Conceptual scheme of the magnetron transmitter available for a fast control in phase and power.

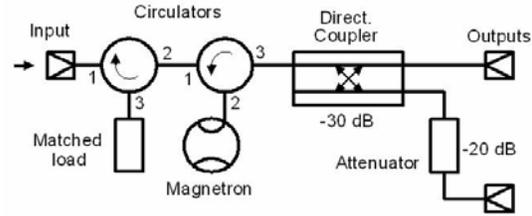

Fig. 2: The magnetron experimental module for tests of the transmitter setup.

The transmitter, Fig. 1, consists of two identical channels (A and B) of cascaded injection-locked magnetrons, combined in power by a 3-dB hybrid. The phase management is provided by controlled simultaneously and equally the phases at inputs of both 2-cascade magnetrons. The power management is provided at a control by phase difference at the inputs of the 2-cascade magnetrons. The 2-cascade injection-locked magnetron system allows reducing the required locking power by 10-15 dB, [5].

The both magnetron modules were fed by a single modulator. The magnetron with lower anode voltage was fed by a compensating divider. Proof of principle of the proposed transmitter concept was demonstrated in experiments at various setups, Figs. 3, 4, measuring phase response of the injection-locked magnetrons without and with phase modulation of the locking signal.

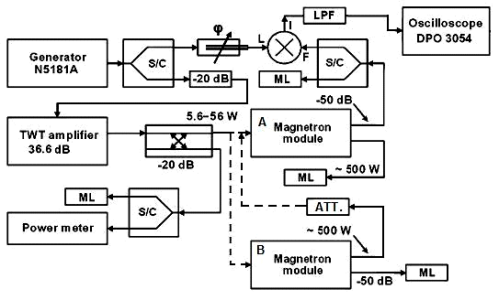

Fig. 3: Setup to test operation of single and 2-cascade injection-locked magnetrons. S/C is a splitter/combiner, LPF is a low pass filter, ML is a matched load, ATT is an attenuator. Powers of the magnetrons are depicted in this figure. The single injection-locked magnetrons were tested in configuration using module A with the magnetron locked by the TWT amplifier and fed by the modulator, while the module B was disconnected from the amplifier and the modulator, [6].

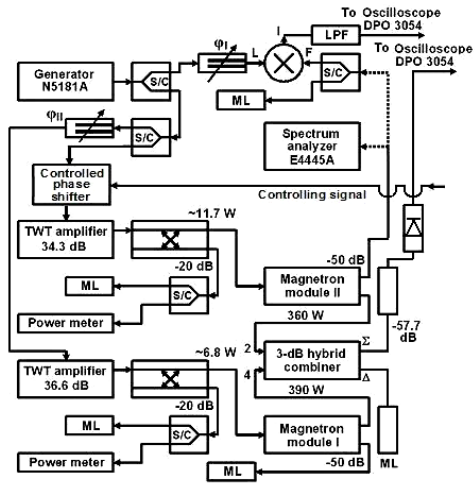

Fig. 4: Setup with the injection-locked magnetrons for test of the power control by power combining, [Ibid.].

Phase behaviour of the magnetrons injection-locked without and with the phase modulation of the locking signal (by phase modulation in the N5181A synthesizer) measured with setups, Figs. 3, 4, is shown in Figs. 5, 6, respectively.

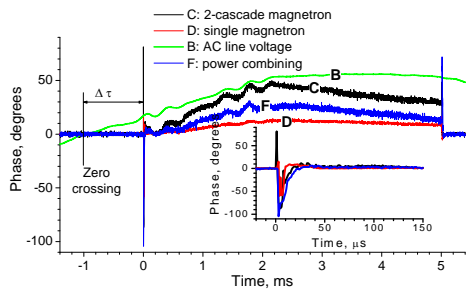

Fig. 5: Phase variations of injection-locked single, 2-cascade magnetron, and magnetrons with power combining, traces D, C, and F, respectively.

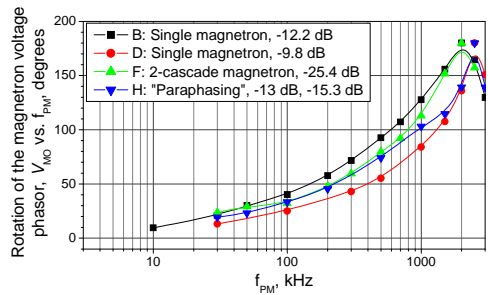

Fig. 6: Angle θ of rotation of the phasor of voltage in the wave at output of the injection-locked magnetrons, caused by transient process of the phase modulation.

The last figure demonstrates measured phase rotation of the voltage phasor in the output wave of the magnetrons for a single tube, 2-cascade setup and magnetrons with power combining vs. the modulating frequency $f_{PM}$, at the phase modulation magnitude of 20 degrees, [7]. The Figs. 5 and 6 verify operation of the considered magnetron setups in injection-locked mode even at wide-band phase modulation ($f_{PM} \leq$ 3 MHz) of the locking signal.

Allowable bandwidth of the phase control of the injection-locking phase-modulated signal was measured at small magnitude of the phase modulation (0.07 rad.) in the phase modulation domain [7]. The magnetron transfer function magnitude characteristic was measured vs. $f_{PM}$, Fig. 7, for single and 2-cascade magnetrons.

The plots in Fig. 8 demonstrate precise stability of carrier frequency of magnetrons, injection-locked by phase-modulated signal. The experiments were conducted in CW mode at wide range of magnitude and frequency modulating the injection-locking signal, [3].

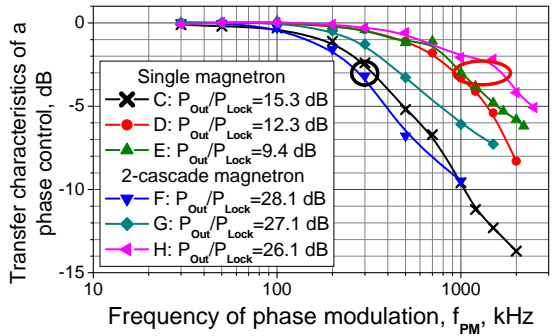

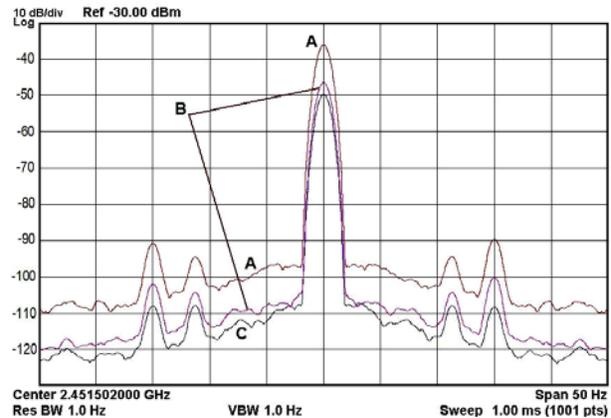

Fig. 7: Transfer function magnitude characteristics (rms values) of the phase control measured in phase modulation domain at various ratios $P_{Out}/P_{Lock}$, at $P_{Out} \approx$ 450 W, [7].

Fig. 8: The carrier frequency spectra of the magnetron injection-locked by a signal without, trace A, and with the phase-modulation. Trace B shows the spectrum of the magnetron injection-locked by the phase-modulated signal at magnitude and frequency of the modulation of 3 radians and 2 MHz, respectively. Trace C shows the carrier frequency spectrum of the locking system (N5181A synthesizer and TWT amplifier).

The 3-dB cutoff clearly indicates that the bandwidth of the phase control is ≥300 kHz at the locking power of ≤-14 dB (per magnetron in 2-cascade setup). Increase of the locking power to ≥-13 dB (per magnetron in 2-cascade setup) increases the bandwidth to over 1 MHz.

The measured transfer characteristics, Fig. 7, imply that a Low Level RF controller with a closed loop at the bandwidth of ≥100 kHz will be able to suppress all expected system low-frequency perturbations, such as parasitic low frequency modulation and phase perturbations from SRF cavity beam loading, the cavity dynamic tuning errors, the low-frequency perturbations caused by magnetron power supplies ripples, etc. For example, the phase pushing in magnetrons caused by ripples ($f_r$=120 Hz) of HV power supply will be suppressed within a phase feedback loop with integral gain of Ig ≈1.2·10$^7$ rad./s by ≈20·log( Ig/2π·$f_r$)≈ 84 dB.

The traces shown in Fig. 8 demonstrate that the carrier frequency of the magnetrons, injection-locked by phase-modulated signal still precisely stable at wide band and wide range of the modulating frequency and magnitude, respectively. It demonstrates proof of principle of the proposed concept of phase-locking in magnetrons by signals, phase-modulated in the wide band. Emulation of the LLRF system with a wideband closed feedback loop demonstrates that all known low-frequency parasitic modulations of accelerating field in SRF cavities will be eliminated; i.e. the injection-locked magnetrons are adequate RF sources for feeding SRF cavities, [3].

Example of application of the wide-band phase modulation in the injection-locked magnetrons with power combining demonstrates proof of principle of the proposed transmitter concept in a wide-range dynamic power control, Fig. 9, [6].

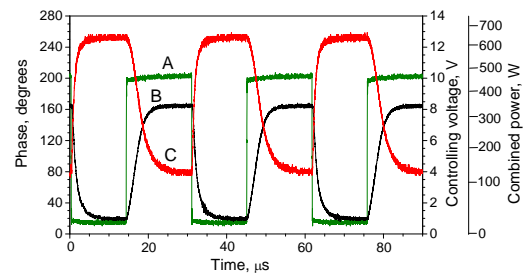

Fig. 9: Trace A shows shape of signals controlling the phase shifter (setup in Fig. 4), first right scale. Trace B is the phase variations at the output of the magnetron II measured by the phase detector, left scale. Trace C shows power measured at port "Σ" of the hybrid combiner at the phase shifter control, second right scale.

**Summary**

Performed R&D demonstrates proof of principle of the proposed concept of the phase wide-band modulation of the injection-locking signal for stabilization of phase and amplitude of the accelerating field in SRF cavities of the intensity-frontier linacs and demonstrates proposed concept of high-power magnetron transmitter with a wide-band phase and power control, intended to feed the SRF cavities.
Results of the conducted R&D substantiate utilization of the proposed transmitter, based on the magnetrons, injection-locked by wide-band phase-modulated signal for powering of SRF cavities to compensate SR losses in the electron ring of MEIC.

**Acknowledgment**

Author cordially thanks Professor Ya. S. Derbenev for setting of scientific target of utilization of the magnetron transmitters for powering of SRF cavities of the MEIC electron ring.